\documentclass[aps,pra,onecolumn,superscriptaddress]{revtex4}
\usepackage{amsfonts,amssymb,graphicx,amsmath}
\usepackage{amsmath}
\usepackage{epstopdf}
\usepackage[usenames,dvipsnames]{xcolor}
\usepackage{pdfpages}
\usepackage{footmisc}
\usepackage[normalem]{ulem}
\usepackage{array}
\usepackage{color}

\begin{document}
\title{Exact solution of the two-magnon problem in the $k=-\pi/2$ sector of a finite-size anisotropic spin-1/2 frustrated ferromagnetic chain}
\author{Zimeng Li}
\affiliation{Center for Quantum Technology Research, and Key Laboratory of Advanced Optoelectronic Quantum Architecture and Measurements (MOE), School of Physics, Beijing Institute of Technology, Beijing 100081, China}
\author{Ning Wu}
\email{wunwyz@gmail.com}
\affiliation{Center for Quantum Technology Research, and Key Laboratory of Advanced Optoelectronic Quantum Architecture and Measurements (MOE), School of Physics, Beijing Institute of Technology, Beijing 100081, China}
\begin{abstract}
The two-magnon problem in the $k=-\pi/2$ sector of a \emph{finite-size} spin-1/2 chain with ferromagnetic nearest-neighbor (NN) interaction $(J_1>0)$ and antiferromagnetic next-nearest-neighbor (NNN) interaction $(J_2<0)$ and anisotropy parameters $\Delta_1$ and $\Delta_2$ is solved exactly by combining a set of exact two-magnon Bloch states and a plane-wave ansatz. Two types of two-magnon bound states (BSs), i.e., the NN and NNN exchange BSs, are revealed. We establish a phase diagram in the $J_1/(|J_2|\Delta_2)$-$\Delta_1$ plane where regions supporting different types of BSs are analytically identified. It is found that no BSs exist (the two types of BSs coexist) when both $\Delta_1$ and $\Delta_2$ are small (large) enough. Our results for the isotropic case $\Delta_1=\Delta_2=1$ are consistent with an early work [Ono I, Mikado S and Oguchi T 1971 \emph{J. Phys. Soc. Japan} \textbf{30} 358].
\end{abstract}
\maketitle
\section{Introduction}
\par Using the ansatz named after him, Bethe showed the existence of two-magnon bound states (BSs) lying above the continuum of the two-magnon band of an antiferromagnetic spin-1/2 XXX chain~\cite{Bethe}. Thereafter few-magnon excitations and BSs in various types of quantum spin systems have become a focus of theoretical investigation. Among these, spin systems with competing interactions are of special interest due to the coexistence of frustration effect and quantum fluctuation. A typical example of such a frustrated system is the $J_1$-$J_2$ chain with ferromagnetic nearest-neighbor (NN) and antiferromagnetic next-nearest-neighbor (NNN) interactions~\cite{Ono,Chubukov,Kecke,Kuzian,Dmiteriev2009,PRB2024_1},
\begin{eqnarray}\label{Haml}
H_{J_1 - J_2}&=& -\sum^2_{\alpha=1} J_\alpha\sum^{N}_{j=1} (S^x_{j}S^x_{j+\alpha}+S^y_{j}S^y_{j+\alpha}+\Delta_\alpha S^z_{j}S^z_{j+\alpha}),
\end{eqnarray}
where $\vec{S}_j=(S^x_j,S^y_j,S^z_j)$ are spin-$S$ operators on site $j$, $J_1>0$ and $J_2<0$ are the exchange interactions between two NN and NNN spins, respectively. The corresponding anisotropy parameters $\Delta_1$ and $\Delta_2$ are assumed to be positive. Although most of the previous works have focused on the isotropic case $\Delta_1=\Delta_2=1$, it is of both theoretical and experimental interest to consider the anisotropic case of the model. On one hand, as shown in both the NN XXZ chain~\cite{Tonegawa,Ganahl2012,PRB2022,PRB2024_2} and the $J_1$-$J_2$ chain~\cite{Kuzian,Dmiteriev2009,PRB2024_1}, finite exchange anisotropy is essential for the formation of few-magnon BSs. On the other hand, recent advances in cold atom physics enable the realization of various types of anisotropic exchange terms~\cite{Fukuhara}. Below we consider an even number of site $N$ and impose periodic boundary conditions $\vec{S}_j=\vec{S}_{N+j}$.
\par Unlike the spin-1/2 NN Heisenberg XXZ chain that can be solved exactly via the Bethe ansatz, the $J_1$-$J_2$ chain is nonintegrable for any quantum number $S$ and its ground-state properties are usually studied by using advanced numerical or field-theoretical methods. Qualitatively, the NNN antiferromagnetic interaction can be viewed as a frustration to the NN ferromagnetic interaction. For an isotropic $J_1$-$J_2$ chain with $S=1/2$ and $\Delta_1=\Delta_2=1$, it is well-known that the ground state is a ferromagnetic state when $J_1/|J_2|\geq 4$~\cite{Kecke,PRB2024_1,Bader1979,Hamada1988,Meisner}. Although the ground state for $J_1/|J_2|< 4$ is no longer ferromagnetic, it is still meaningful to study few-spin deviations from the fully polarized state. Technically, both the spin-wave theory and the Bethe ansatz method start with considerations of high magnetization sectors having a few number of down spins. In the context of the frustrated ferromagnetic chain, Ono, Mikado, and Oguchi use the Jordan-Wigner transformation to study the two-magnon BSs in an infinite isotropic spin-1/2 $J_1$-$J_2$ chain and obtained analytical solutions in the two subspaces with wave numbers $k=-\pi$ and $k=-\pi/2$~\cite{Ono}. Chubukov proposes to study the instability of the ferromagnetic ground state by investigating the two-magnon bound-state spectrum~\cite{Chubukov}. Kecke, Momoi, and Furusaki calculated the $n$-magnon excitation spectra ($n\leq 4$) from the ferromagnetic state in the presence of a saturation field by combining few-magnon Bloch states and a Hilbert space truncation~\cite{Kecke}. Kuzian and Drechsler calculate the two-magnon excitation spectrum of an infinite frustrated spin-1/2 XXZ chain by mapping the two-magnon problem onto a tight-binding impurity problem~\cite{Kuzian}. Dmitriev and Krivnov study the excitation spectrum of a spin-1/2 $J_1$-$J_2$ chain by combining perturbation theory and numerical calculations~\cite{Dmiteriev2009}. Recently, Li, Cao, and Wu generalize the two-magnon problem to a higher-spin anisotropic $J_1$-$J_2$ chain with a single-ion anisotropy term~\cite{PRB2024_1} and use a set of exact two-magnon Bloch states~\cite{PRB2022} to obtain the full two-magnon spectrum for finite-size chains. In particular, the problem in the $k=-\pi$ subspace is solved analytically by using a plane-wave ansatz, which allows one to quantitatively understand the emergence of different types of two-magnon BSs near the edge of the Brillouin zone~\cite{PRB2024_1}.
\par In this work, we combine the Bloch-state method with the plane-wave ansatz to exactly solve the two-magnon problem in the $k=-\pi/2$ sector of a finite-size anisotropic spin-1/2 $J_1$-$J_2$ chain. We find that the formation of two-magnon BSs in this mode depends on the two parameters, i.e., the NN anisotropy $\Delta_1$ and the ratio $J_1/(|J_2|\Delta_2)$. We analytically reveal in the $J_1/(|J_2|\Delta_2)$-$\Delta_1$ plane several regions that support different types of two-magnon BSs, where the boundaries are defined by algebraic equations. As expected, the parameter $\Delta_1$ or $J_1/(|J_2|\Delta_2)$ roughly determines the formation of the so-called nearest-neighbor exchange (NN-Ex) or next-nearest-neighbor exchange (NNN-Ex) BS, where the two down spins mainly locate on two NN or NNN sites. From physical continuity, we expect that these analytical results can faithfully reflect the emergence of two-magnon BSs in the middle of the half-Brillouin zone.
\par The rest of the paper is organized as follows. In Sec.~\ref{SecII} we introduce the dimensionless $J_1$-$J_2$ chain and the methods to deal with the two-magnon problem, i.e., the exact two-magnon Bloch states and the plane-wave ansatz. We derive a transcendental equation that can determine the explicit forms of the eigenenergies and wave functions of the two-magnon eigenstates. In Sec.~\ref{SecIII} we study in detail the complex solutions of the transcendental equation and analytically determine the properties of the two-magnon BSs in different parameter regions. Conclusions are drawn in Sec.~\ref{SecVI}.
section{Model and methodology}\label{SecII}
\subsection{Model and the Bloch Hamiltonian for $k=-\pi/2$}\label{SecII1}
\par We consider a dimensionless spin-1/2 ferromagnetic-antiferromagnetic chain described by the Hamiltonian $H\equiv H_{J_1 - J_2}/|J_2|$,
\begin{eqnarray}\label{Haml}
H&=& - \gamma\sum^{N}_{j=1} (S^x_{j}S^x_{j+1}+S^y_{j}S^y_{j+1}+\Delta_1 S^z_{j}S^z_{j+1})\nonumber\\
 &&+ \sum^{N}_{j=1} (S^x_{j}S^x_{j+2}+S^y_{j}S^y_{j+2}+\Delta_2 S^z_{j}S^z_{j+2}),
\end{eqnarray}
where the positive dimensionless parameter $\gamma\equiv J_1/|J_2|$ measures the ferromagnetic exchange interaction between two NN spins. The total magnetization $M=\sum^N_{j=1}S^z_j$ is conserved and the fully polarized state $|F\rangle=|\uparrow\uparrow\cdots\uparrow\rangle$ is an obvious eigenstate of $H$ with  eigenenergy $E_F=-N(\gamma\Delta_1 -\Delta_2)/4$. We take $|F\rangle$ as a reference state and study the two-magnon excitations upon this state~\cite{Ono}. Note that $|F\rangle$ is not necessarily a ground state of $H$. The $N(N-1)/2$ real-space two-magnon basis states are given by $|i,j\rangle\equiv S^-_iS^-_j|F\rangle$ ($i<j$), where $S^-_l=S^x_l-iS^y_l$ is the spin lowering operator on site $l$. Following Ref.~\cite{PRB2022}, we introduce the following two-magnon Bloch states for a finite-size chain, 
\begin{eqnarray}\label{Bloch2m1}
|\xi_r(k)\rangle&=&\frac{e^{i\frac{rk}{2}}}{\sqrt{N}}\sum^{N-1}_{n=0}e^{ikn}T^{n}|1,1+r\rangle,~1\leq r\leq \frac{N}{2}-1
\end{eqnarray}
where $k\in K_0=\left\{-\pi,-\pi+\frac{2\pi}{N},\cdots,0,\cdots,\pi-\frac{2\pi}{N}\right\}$, and
\begin{eqnarray}\label{Bloch2m2}
|\xi_{\frac{N}{2}}(k)\rangle&=&e^{i\frac{Nk}{4}}\sqrt{\frac{2}{N}}\sum^{\frac{N}{2}-1}_{n=0}e^{ikn}T^{n}|1,1+\frac{N}{2}\rangle,
\end{eqnarray}
where $k\in K_1=\left\{-\pi,-\pi+\frac{4\pi}{N},\cdots,0,\cdots,\pi-\frac{4\pi}{N}\right\}$ (for even $N/2$). Here, $T$ is the translation operator defined by $T|i,j\rangle=|i+1,j+1\rangle$ and the index $r$ in $|\xi_r(k)\rangle$ measures the distance between the two down spins on the ring. Since $-\pi/2\notin K_0$ when $N=8m+2$ or $N=8m+6$ ($m$ is a positive integer), we always assume that $N$ is divisible by 4. The complement of $K_1$ is denoted as $K'_1$ (i.e., $K_0=K_1\bigcap K'_1$)~\cite{PRB2024_1,PRB2022,PRB2024_2}, then $-\pi/2\in K_1$ if $N=8m$ and $-\pi/2\in K'_1$ if $N=8m+4$. Note that $|\xi_{\frac{N}{2}}(k)\rangle$ is not defined for $k\in K'_1$. In this work, we focus on the case of $N=8m$.
\begin{figure}
\begin{center}
\includegraphics[width=.5\textwidth]{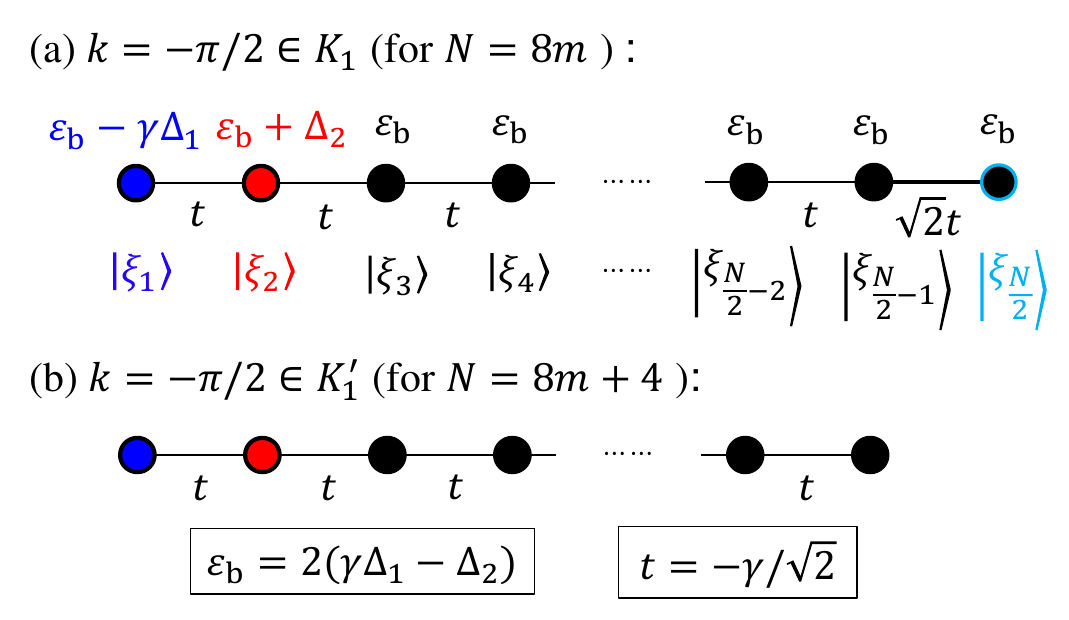}
\caption{(a) Matrix form of $H-E_F$ in the ordered basis $\{|\xi_1(-\pi/2),\ldots,\xi_{N/2}(-\pi/2)\}$ with $k=-\pi/2\in K_1$ when $N=8m$. It corresponds to an inhomogeneous open chain having nearest-neighbor hopping.  (b) The case of $k=-\pi/2\in K'_1$ when $N=8m+4$.}
\label{Fig1}
\end{center}
\end{figure}
\begin{figure}
\begin{center}
\includegraphics[width=.8\textwidth]{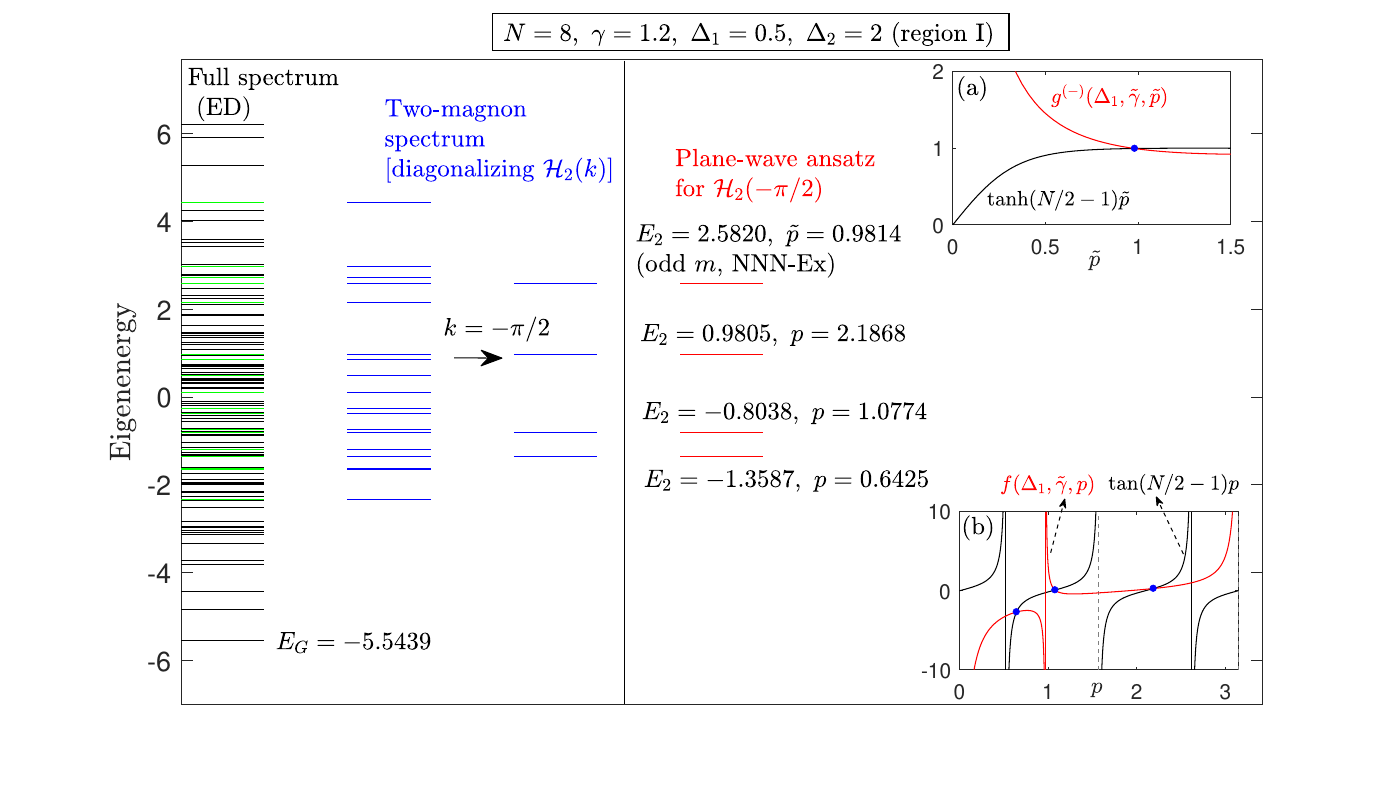}
\caption{Left panel: Full spectrum of $H_{J_1-J_2}$ (black) for $N=8,~\gamma=1.2,~\Delta_1=0.5$, and $\Delta_2=2$. The ground-state energy is $E_G=-5.5439$ and the two-magnon spectrum in the $M=2$ sector (consisting of 18 distinct levels) is indicated in green. The blue lines are the two-magnon spectrum  calculated by diagonalizing the Bloch Hamiltonians $\mathcal{H}_2(k)$, where the four eigenenergies in the $k=-\pi/2$ sector are shown separately. Right panel: The four eigenenergies calculated by using the plane-wave ansatz solution of $\mathcal{H}_2(-\pi/2)$. Inset (a) shows the graphic solution of Eq.~(\ref{tanh_gm}), giving $\tilde{p}=0.9814$; while inset (b) shows the graphic solutions of Eq.~(\ref{realsolEQ}), giving three solutions $p=2.1868$, $1.0774$, and $0.6425$. The resulting eigenenegies are respectively $E_2=2.5820$, $0.9805$, $-0.8038$, and $-1.3587$, which are consistent with the exact diagonalization results.}
\label{check}
\end{center}
\end{figure}
\par As detailed in Ref.~\cite{PRB2024_1}, for a fixed $k\in K_1$ ($k\in K'_1$) the matrix representation of the Hamiltonian $H$ in the ordered basis $\{|\xi_{1\leq r\leq N/2}(k\in K_1)\}$ [$\{|\xi_{1\leq r<N/2}(k\in K'_1)\}$] is a $N/2$-dimensional [$(N/2-1)$-dimensional] pentadiagonal block matrix $\mathcal{H}_2(k\in K_1)$ [$\mathcal{H}_2(k\in K'_1)$], which corresponds to a single-particle problem on a $N/2$-site  [$(N/2-1)$-site] open chain with both NN and NNN hoppings and inhomogeneous onsite energies (see Fig.~\ref{Fig1} for the case of $k=-\pi/2$). To verify the validity of the above block-diagonalization procedure, we numerically diagonalize the matrices $\{\mathcal{H}_2(k\in K_1)\}$ and $\{\mathcal{H}_2(k\in K'_1)\}$ to obtain the whole two-magnon spectrum. In the left panel of Fig.~\ref{check}, we present the full energy spectrum of $H_{J_1-J_2}$ (leftmost) calculated by exact diagonalization (ED) of a small ring with $N=8$, $\gamma=1.2$, $\Delta_1=0.5$, and $\Delta_2=2$. The ground-state energy is found to be $E_G=-5.5439$, which lies well below the two-magnon spectrum (green), indicating that the ground state is not in the two-magnon sector. The two-magnon spectrum calculated by numerically diagonalizing the Bloch Hamiltonians is shown on the right (blue) and is consistent with the ED results (green). Due to the reflection symmetry of the chain, there are only 18 distinct energy levels in the 28-dimensional two-magnon subspace. The four energy levels for $k=-\pi/2$ are also selected out and take values $E_2=-1.2587$, $-0.8038$, $0.9805$, and $2.5820$. 
\par Although the numerical diagonalization of the Bloch Hamiltonians is straightforward and can give accurate eigenenergies of the system, it is more desirable to analytically or semianalytically solve these Hamiltonians so as to get closed-form expressions for the wave functions and eigenenergies. Unfortunately, the block matrices $\mathcal{H}_2(k)$ generally cannot be solved analytically, except for a few special cases. For example, it is shown in Ref.~\cite{PRB2024_1} that for the particular mode $k=-\pi$ the NN hopping within the effective open chain vanishes and one gets two decoupled NN chains that can be solved semianalytically by using a plane-wave ansatz, which is a standard method applicable to several inhomogeneous problems~\cite{Grim,PRB2017,PRB2019}. The key to the plane-wave ansatz lies in the fact that the effective Hamiltonians admit the form of the following inhomogeneous tridiagonal matrix~\cite{PRB2024_1,PRB2024_2}:
\begin{eqnarray}\label{Mmatrix}
O=\left(
      \begin{array}{cccccccc}
        a_1 &  b_1 &   & &    &    &   &   \\
         b_1 & a_2 &  b &  &   &   &   &   \\
          &  b & 0 &  b  & &   &   &   \\
          &   &  b &  0 &  &    &   &   \\
           &   &    &  & \ddots&   &   &   \\
          &   &   &     & &0 &  b &   \\
          &   &   &    &  &b & 0 &   b_2 \\
          &   &   &    & &  &  b_2  & a_3 \\
      \end{array}
    \right).
\end{eqnarray} 
\par We now note that for $k=-\pi/2$ the NNN hopping in the open chain is zero, yielding an inhomogeneous NN chain [in the ordered basis $\{|\xi_1(-\pi/2),\ldots,\xi_{N/2}(-\pi/2)\}$] as shown in Fig.~\ref{Fig1}, which can be incorporated into the $O$ matrix via the correspondence
\begin{eqnarray} 
a_1=-\gamma\Delta_1,~a_2=\Delta_2,~a_3=0,~b=b_1=t,~b_2=\sqrt{2}t.
\end{eqnarray} 
It is worth emphasizing that for higher-spin $J_1$-$J_2$ chains with $S>1/2$ the effective NN chain has an extra site (defined by $|\xi_0(-\pi/2)\rangle$) as its leftmost site, so that the corresponding Bloch Hamiltonian cannot fit into the $O$ matrix since the first three diagonal elements are in general distinct. Consequently, to get semianalytical solutions of the problem, we are forced to consider the case of $S=1/2$ in this work. 
\par
\begin{figure}
\begin{center}
\includegraphics[width=.7\textwidth]{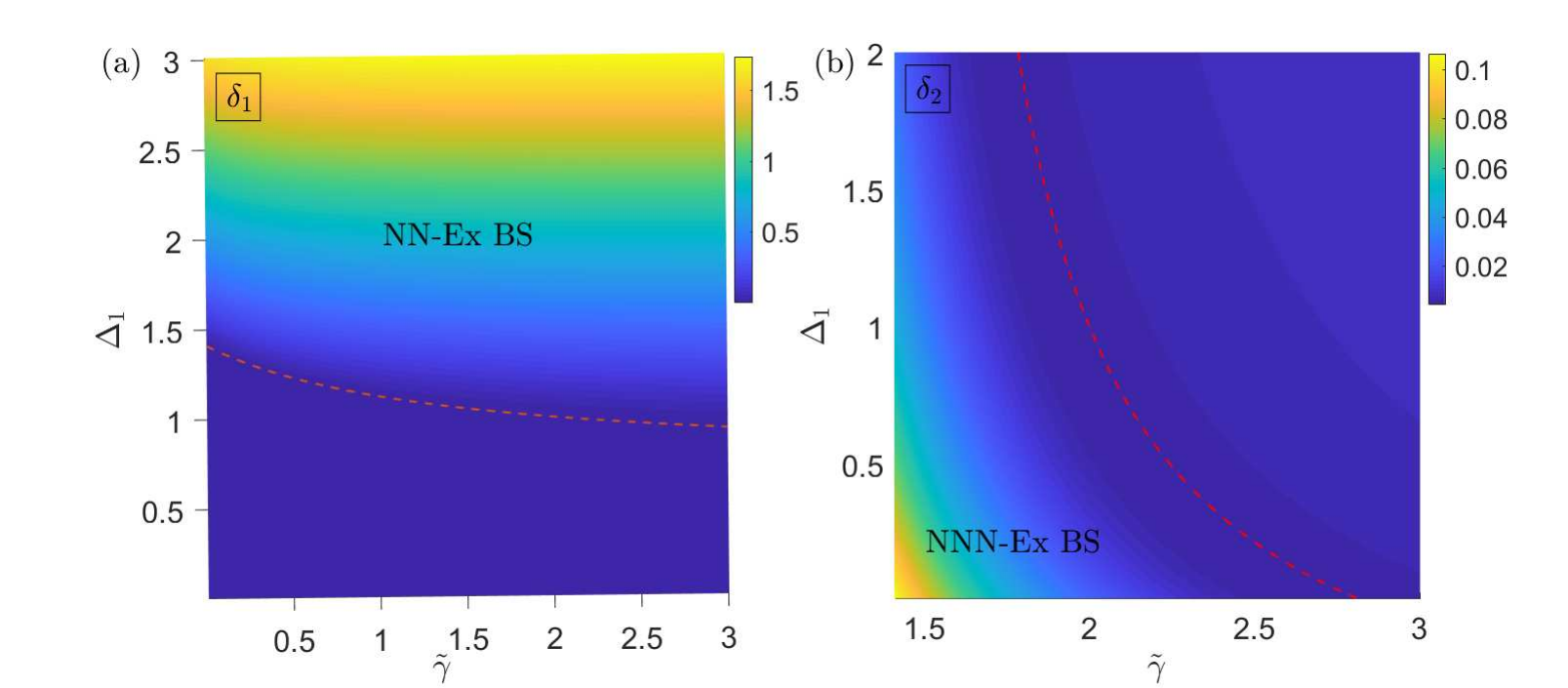}
\caption{The energy differences (a) $\delta_1= E^{(2)}_2-E^{(1)}_2$ and (b) $\delta_2= E^{(N/2)}_2-E^{(N/2-1)}_2$ for $N=80$ and $\gamma=1$. The red dashed curves roughly indicate the boundary where the two-magnon bound states start to appear.}
\label{Figdiff}
\end{center}
\end{figure}

 The inhomogeneity of the chain lies in the onsite energies of the first two sites $|\xi_1(-\pi/2)\rangle$ and $|\xi_2(-\pi/2)\rangle$, as well as the hopping between the two sites $|\xi_{N/2-1}(-\pi/2)\rangle$ and $|\xi_{N/2}(-\pi/2)\rangle$ when $N=8m$. From Fig.~\ref{Fig1}(a) we read off the matrix representation of $H$ as $\mathcal{H}_2(-\pi/2)=E_F+\varepsilon_{\mathrm{b}}-\gamma h$, where $h$ is a $\frac{N}{2}\times\frac{N}{2}$ matrix 
\begin{eqnarray}\label{hnn}
h&=&\left(
      \begin{array}{cccccccc}
          \Delta_1 &   \frac{1}{\sqrt{2}} &   & &    &    &   &   \\
           \frac{1}{\sqrt{2}} & -1/\tilde{\gamma} &   \frac{1}{\sqrt{2}} &  &   &   &   &   \\
          &   \frac{1}{\sqrt{2}} & 0 &   \frac{1}{\sqrt{2}}  & &   &   &   \\
          &   &   \frac{1}{\sqrt{2}} &  0 &  &    &   &   \\
           &   &    &  & \ddots&   &   &   \\
          &   &   &     & &0 &   \frac{1}{\sqrt{2}} &   \\
          &   &   &    &  & \frac{1}{\sqrt{2}} & 0 &   1 \\
          &   &   &    & &  &   1  & 0 \\
      \end{array}
    \right),
\end{eqnarray}
with $\varepsilon_{\mathrm{b}}\equiv 2(\gamma\Delta_1-\Delta_2)$ and $\tilde{\gamma}\equiv\gamma/\Delta_2$. The structure of the eigenstates of $\mathcal{H}_2(-\pi/2)$ depends solely on the matrix $h$ via the two parameters $\Delta_1$ and $\tilde{\gamma}$. By solving the eigenproblem $\mathcal{H}_2(-\pi/2)V^{(\alpha)}=E^{(\alpha)}_2V^{(\alpha)}$~($\alpha=1,2,\ldots,N/2$), we can obtain the two-magnon spectrum $E^{(1)}_2\leq E^{(2)}_2<\cdots\leq E^{(N/2)}_2$ in the $k=-\pi/2$ sector. We are mainly interested in the formation of two-magnon BSs, whose energies usually separate from the scattering continuum~\cite{PRB2024_1}. An NN-Ex (NNN-Ex) two-magnon BS is defined via the relation $|V^{(\alpha)}_1|>|V^{(\alpha)}_2|>|V^{(\alpha)}_3|>\cdots>|V^{(\alpha)}_{N/2}|$ ($|V^{(\alpha)}_1|<|V^{(\alpha)}_2|>|V^{(\alpha)}_3|>\cdots>|V^{(\alpha)}_{N/2}|$), which corresponds to an eigenstate with the two down spins mainly locating on the two NN (NNN) sites~\cite{PRB2024_1}. Intuitively, an NN-Ex (NNN-Ex) BS is expected to lie below (above) the continuum due to the ferromagnetic (antiferromagnetic) nature of the NN (NNN) exchange interaction~\cite{PRB2024_1} since the NN Ising coupling $-\gamma\Delta_1S^z_jS^z_{j+1}$ (NNN Ising coupling $\Delta_2S^z_jS^z_{j+2}$ ) dominates for large enough $\Delta_1$ ($\Delta_2$). Actually, as we will show rigorously later, an NN-Ex (NNN-Ex) BS (if it exists) is given by $V^{(1)}$ ($V^{(N/2)}$) and indeed lies below (above) the continuum. As a result, the magnitude of the energy difference between the second lowest and the lowest eigenenergies $\delta_1\equiv E^{(2)}_2-E^{(1)}_2 $ [the highest and the second highest eigenenergies $\delta_2\equiv E^{(N/2)}_2-E^{(N/2-1)}_2 $] may roughly serve as an indicator of the existence of an NN-Ex (NNN-Ex) BS. In Fig.~\ref{Figdiff} we plot both $\delta_1$ and $\delta_2$ in the $\tilde{\gamma}$-$\Delta_1$ plane (we choose $N=80$ and $\gamma=1$). The red curve in Fig.~\ref{Figdiff}(a) [Fig.~\ref{Figdiff}(b)] roughly separates parameter regions that support an NN-Ex (NNN-Ex) BS or not. Our aim is to analytically determine the equations for these boundary curves in the $\tilde{\gamma}$-$\Delta_1$ plane. 
\subsection{Plane-wave ansatz solution of the Bloch Hamiltonian}
\par Instead of numerically diagonalizing the Bloch Hamiltonian, here we use a plane-wave ansatz as a trial eigenvector of $\mathcal{H}_2(-\pi/2)$ or $h$,
\begin{eqnarray}\label{VjXY}
V_j=X e^{ipj}+Ye^{-ipj},~j=2,3,\ldots,N/2-1
\end{eqnarray}
where $X$ and $Y$ are $j$-independent coefficients and $p$ is a pseudo-momentum to be determined. Inserting the ansatz into Eq.~(\ref{hnn}) gives the eigenenergy 
\begin{eqnarray}\label{E2}
E_2=E_F+\varepsilon_{\mathrm{b}}-\sqrt{2}\gamma \cos p,
\end{eqnarray}
and the two end components
\begin{eqnarray}\label{endcomp}
V_1=\frac{1 }{ 2 \cos p- \sqrt{2}\Delta_1}V_2,~~V_{N/2}=\frac{1}{ \sqrt{2}  \cos p}V_{N/2-1}.
\end{eqnarray} 
Following Ref.~\cite{PRB2024_1}, we obtain after a straightforward calculation the following transcendental equation that determines the allowed values of $p$:
\begin{eqnarray}\label{realsolEQ}
\tan\left(\frac{N}{2}-1\right)p&=&  f(\Delta_1,\tilde{\gamma},p),\nonumber\\
f(\Delta_1,\tilde{\gamma},p)&\equiv& \frac{ w^{(+)}(\cos p)}{(A- \cos p)\sin p},
\end{eqnarray}
where
\begin{eqnarray}
\tilde{\gamma}&\equiv&  \gamma/ \Delta_2,\nonumber\\
A&\equiv&  \sqrt{2}\tilde{\gamma}/4+  \Delta_1 /\sqrt{2}>0,\nonumber\\
w^{(\pm)}(x)&\equiv& x^2\pm (A-\sqrt{2}\Delta_1)x-\tilde{\gamma}\Delta_1/2.
\end{eqnarray}
Note that $w^{(\pm)}(x)=w^{(\mp)}(-x)$. For each allowed $p$, the (unnormalized) wave function is given by~\cite{PRB2024_1}
\begin{eqnarray}\label{wfVj}
V_j=\cos[(j-N/2)p].
\end{eqnarray}
\begin{table*}\label{TableI}
\begin{center}
\begin{tabular}{|c| c| c| c| c|}
 \hline
 Region & Signs & Number of real sols & $\lim_{\tilde{p}\to 0^+}g^{(\pm)}(\Delta_1,\tilde{\gamma},\tilde{p})$  & Bound states   \\
 \hline
 I & ($-,+,+$) & $N/2-1$ & $(+,+)\infty$  & NNN-Ex \\
 \hline
 II & ($+,+,+$) & $N/2-1$ & $(-,+)\infty$  &  NNN-Ex   \\
 \hline
 III & ($+,-,+$) & $N/2-2$ & $(+,+)\infty$  &  NNN-Ex, NN-Ex  \\
 \hline
 IV & ($+,-,-$) & $N/2-1$ &  $(+,-)\infty$ & NN-Ex \\
 \hline
  V & ($+,+,-$) & $N/2$ &  $(-,-)\infty$  & No \\
 \hline
\end{tabular}
\caption{The signs of $(A-1,w^{(+)}(1),w^{(+)}(-1))$, the number of real solutions of Eq.~(\ref{realsolEQ}), the limits $\lim_{\tilde{p}\to 0^+}g^{(\pm)}(\Delta_1,\tilde{\gamma},\tilde{p})$, and the corresponding two-magnon bound states in regions I-V. Note that the first three columns are directly related: the sign of $w^{(+)}(1)$ [$w^{(+)}(-1)$] determines the existence or absence of a real solution of Eq.~(\ref{realsolEQ}) in the first (last) interval $[0,\pi/(N-2)]$ ($[(N-3)\pi/(N-2),\pi]$), while when $A-1<0$ there exist two solutions in the interval containing $\arccos A$.}
\end{center}
\end{table*}
\par Equation (\ref{realsolEQ}) may have either real or complex solutions. The real solutions of Eq.~(\ref{realsolEQ}) correspond to the scattering states having oscillating wave functions and should be pursued on the interval $p\in[0,\pi]$. The function $\tan(N/2-1)p$ diverges at $p=\pi/(N-2),3\pi/(N-2),\ldots,(N-3)\pi/(N-2)$, which divide the interval $[0,\pi]$ into $N/2$ ones: $[0,\pi/(N-2)],[\pi/(N-2),3\pi/(N-2)],~\ldots,[(N-3)\pi/(N-2),\pi]$. Note that $\tan(N/2-1)p\geq 0$ ($\leq 0$) on the first (last) interval. Therefore, the number of real solutions depends on the singular behaviors of the function $f(\Delta_1,\tilde{\gamma},p)$ as $p\to 0^+$, $p\to \pi^-$, and $\cos p\to A^{\pm}$ (if $A<1$)~\cite{PRB2024_2}. As an example, let us look at the set of parameters used in Fig.~(\ref{check}), for which we have $\tilde{\gamma}=0.6$, $A=2\sqrt{2}/5<1$, $w^{(+)}(1)=(17-2\sqrt{2})/20>0$, and $w^{(+)}(-1)=(17+2\sqrt{2})/20>0$, giving $\lim_{p\to 0^+}f(\Delta_1,\tilde{\gamma},p)=-\infty$ and $\lim_{p\to \pi^-}f(\Delta_1,\tilde{\gamma},p)=+\infty$ [see inset (b) in Fig.~\ref{check}]. As a result, there are no intersections of $\tan 3p$ and $f(\Delta_1,\tilde{\gamma},p)$ on the two end intervals $[0,\pi/6]$ and $[5\pi/6,\pi]$. However, the condition $A<1$ leads to two intersections on the interval containing $\arccos A$, i.e., $[\pi/6,\pi/2]$. Therefore, Eq.~(\ref{realsolEQ}) has three real solutions and a single complex solution. Similar analysis can be done for other combinations of the signs of the tuple $(A-1,w^{(+)}(1),w^{(+)}(-1))$. The equations $A-1=0$, $w^{(+)}(1)=0$, and $w^{(+)}(-1)=0$ respectively define the curves $\Delta_1=-\frac{1}{2}\tilde{\gamma}+\sqrt{2}$, $\Delta_1=\frac{4+ \sqrt{2}\tilde{\gamma} }{2\sqrt{2}+2\tilde{\gamma} }$, and $\Delta_1= \frac{4-\sqrt{2}\tilde{\gamma} }{-2\sqrt{2}+2\tilde{\gamma} }$ in the $\tilde{\gamma}$-$\Delta_1$ plane, which divide the first quadrant of the $\tilde{\gamma}$-$\Delta_1$ plane into five regions I, II, III, IV, and V (see Fig.~\ref{Fig2}). As we will see, the latter two curves (blue and pink curves in Fig.~\ref{Fig2}) just correspond to the two red dashed curves in Fig.~\ref{Figdiff}. 
\par The signs of $(A-1,w^{(+)}(1),w^{(+)}(-1))$ in each region are listed in the first column of Table~1, from which we obtain the corresponding number of real solutions of Eq.~(\ref{realsolEQ}) (see the second column of Table~1). Accordingly, there are $1,~1,~2,~1$, and zero complex solutions in regions I, II, III, IV, and V, respectively. The complex solutions of Eq.~(\ref{realsolEQ}) correspond to two-magnon BSs and will be studied in detail in the next section.
\section{Complex solutions: Two-magnon bound states}\label{SecIII}
\begin{figure} 
\begin{center}
\includegraphics[width=.5\textwidth]{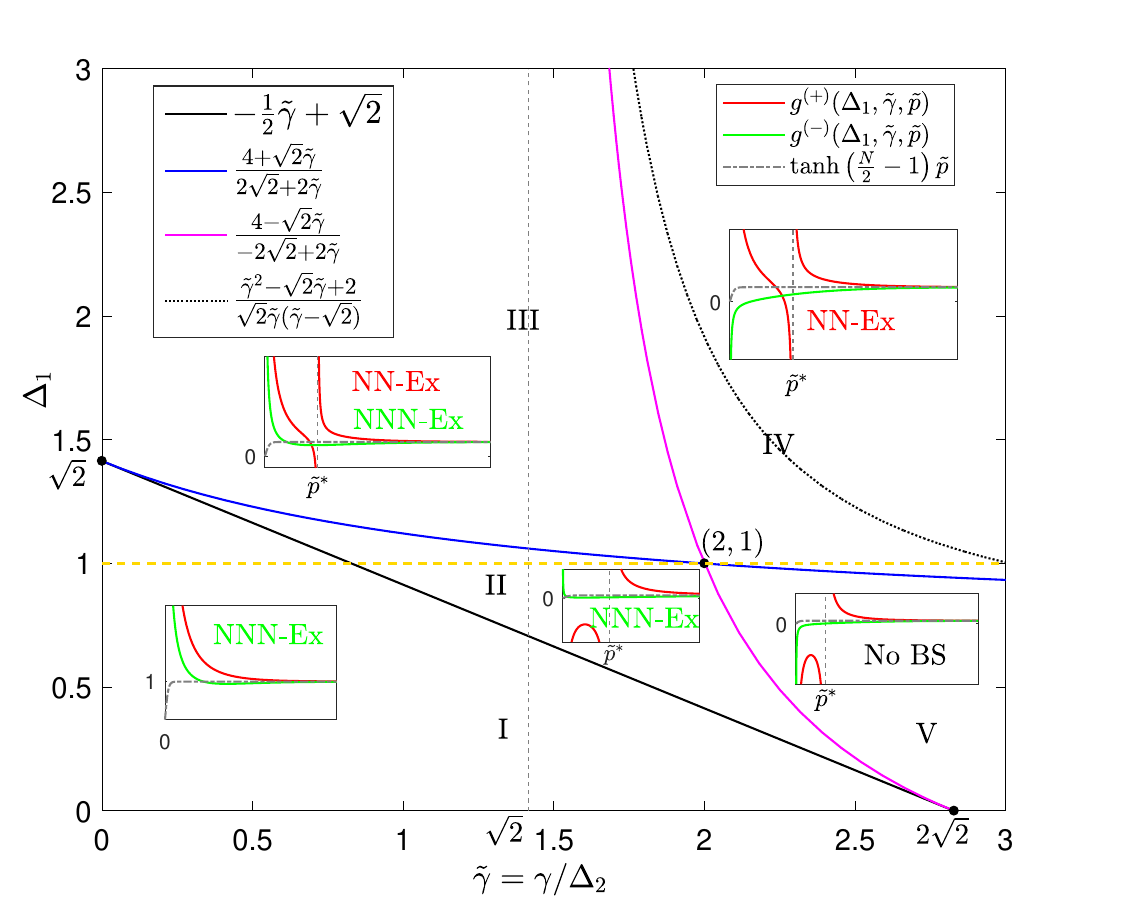}
\caption{The first quadrant of the $\tilde{\gamma}$-$\Delta_1$ plane is divided into five regions I, II, III, IV, and V by the three functions $\Delta_1=-\frac{1}{2}\tilde{\gamma}+\sqrt{2}$ (black), $\Delta_1=\frac{4+ \sqrt{2}\tilde{\gamma} }{2\sqrt{2}+2\tilde{\gamma} }$ (blue), and $\Delta_1= \frac{4-\sqrt{2}\tilde{\gamma} }{-2\sqrt{2}+2\tilde{\gamma} }$ (pink). Insets: The intersections of the graphs of $\tanh\left(\frac{N}{2}-1\right)\tilde{p}$ and $g^{(\pm)}(\Delta_1,\tilde{\gamma},\tilde{p})$ give complex solutions of Eq.~(\ref{realsolEQ}) or real solutions of Eqs.~(\ref{tanh_gp}) and (\ref{tanh_gm}). The dashed orange line defined by $\Delta_1=1$ indicates the evolution of the BS with varying $\gamma$ in the isotropic case~\cite{Ono}.}
\label{Fig2}
\end{center}
\end{figure}
\par We now pursue complex solutions of Eq.~(\ref{realsolEQ}) in regions I-IV. To ensure the reality of the eigenenergy $E_2$, any complex solution must have the form $p=m\pi+i\tilde{p}$~\cite{PRB2024_1,PRB2017,PRB2019}, where $m\in\mathbb{Z}$ and $\tilde{p}\geq 0$. Using the relations $\cos p=(-1)^m\cosh\tilde{p},~\sin p=i(-1)^m\sin\tilde{p}$, we get
\begin{eqnarray}\label{tanh_gp}
\tanh\left(\frac{N}{2}-1\right)\tilde{p}&=& g^{(+)}(\Delta_1,\tilde{\gamma},\tilde{p}),\nonumber\\
 g^{(+)}(\Delta_1,\tilde{\gamma},\tilde{p})&\equiv&-\frac{ w^{(+)}(\cosh \tilde{p})}{(A-\cosh \tilde{p})\sinh \tilde{p}}
\end{eqnarray}
for even $m$, and
\begin{eqnarray}\label{tanh_gm}
\tanh\left(\frac{N}{2}-1\right)\tilde{p}&=& g^{(-)}(\Delta_1,\tilde{\gamma},\tilde{p}),\nonumber\\
 g^{(-)}(\Delta_1,\tilde{\gamma},\tilde{p})&\equiv&\frac{ w^{(-)}(\cosh \tilde{p})}{(A+\cosh \tilde{p})\sinh \tilde{p}}
\end{eqnarray}
for odd $m$.
\par It is easy to see that
\begin{eqnarray}\label{limit1}
\lim_{\tilde{p}\to\infty}g^{(\pm)}(\Delta_1,\tilde{\gamma},\tilde{p})=1^{\pm}.
\end{eqnarray}
For $A>1$ (regions II-V), the function $ g^{(+)}(\Delta_1,\tilde{\gamma},\tilde{p})$ is singular at $\tilde{p}=\tilde{p}^*\equiv\cosh^{-1}A$. From $w^{(+)}(A)=  \tilde{\gamma}^2/4>0$, we have
\begin{eqnarray}\label{limit2}
\lim_{\tilde{p}\to \tilde{p}^{*\pm}}g^{(+)}(\Delta_1,\tilde{\gamma},\tilde{p}) =\pm\infty.
\end{eqnarray}
In addition, the limits of $g^{(\pm)}(\Delta_1,\tilde{\gamma},\tilde{p})$ as $\tilde{p}\to 0^+$ are summarized in the fourth column of Table~1. These limiting behaviors of $g^{(+)}(\Delta_1,\tilde{\gamma},\tilde{p})$ are shown in the insets of Fig.~\ref{Fig2} and essentially determine the properties of real solutions of Eqs.~(\ref{tanh_gp}) and (\ref{tanh_gm}) in each region. As expected, Eq.~(\ref{tanh_gp}) has one real solution in regions III and IV on the interval $[0,\tilde{p}^*]$ and Eq.~(\ref{tanh_gm}) has one real solution in regions I, II, and III on $[0,\infty)$.
\par We will show that the solution of Eq.~(\ref{tanh_gp}) [Eq.~(\ref{tanh_gm})] corresponds to an NN-Ex (NNN-Ex) two-magnon BS. As mentioned in Sec.~\ref{SecII1}, the nature of a BS can be seen from the relative position of its energy level with respect to the scattering continuum. From Eq.~(\ref{E2}) we have
\begin{eqnarray}
E_2=E_F+\varepsilon_{\mathrm{b}}+(-1)^{m-1} \sqrt{2}\gamma \cosh\tilde{p}.
\end{eqnarray}
We see that $E_2$ indeed lies above (below) the continuum if $m$ is odd (even) since $\cosh\tilde{p}\geq \cos p$. For completeness, below we will reconfirm this by investigating the wave function of a BS.
\par For $j=2,\ldots,\frac{N}{2}-1$, we obtain from Eq.~(\ref{wfVj}) the bulk components of the wave function,
\begin{eqnarray}
V_j=(-1)^{mj}\cosh[(N/2-j)\tilde{p}].
\end{eqnarray}
The two end components are given by Eq.~(\ref{endcomp}) as
\begin{eqnarray}
V_1&=& RV_2,~V_{N/2}=\frac{(-1)^m }{ \sqrt{2}  \cosh\tilde{p} }V_{N/2-1},
\end{eqnarray}
where
\begin{eqnarray}
R\equiv \frac{1 }{(-1)^{m}2 \cosh\tilde{p}- \sqrt{2}\Delta_1}.
\end{eqnarray}
It is easy to see that $|V_2|<|V_3|<\cdots<|V_{N/2-1}|<|V_{N/2}|$. Thus, the state is an NN-Ex (NNN-Ex) BS if $|R|>1$ ($|R|<1$).
\par We now show that for even $m$ we always have $R>1$ in regions III and IV, reconfirming the fact that an NN-Ex BS exists in these two regions. To this end, we consider the large $N$ limit~\cite{PRB2024_2}. By using the approximation $\tanh\left(N/2-1\right)\tilde{p}\approx 1$ and squaring both sides of Eq.~(\ref{tanh_gp}), we arrive at the following cubic equation satisfied by $x\equiv \cosh\tilde{p}$, 
\begin{eqnarray}\label{Fx0}
F(x)=0,
\end{eqnarray}
where
\begin{eqnarray}\label{Fx}
F(x)&\equiv&8\sqrt{2}\tilde{\gamma} x^3+8(1-2\Delta_1\tilde{\gamma})x^2-2\sqrt{2}(4\Delta_1+2\tilde{\gamma}-2\Delta^2_1\tilde{\gamma}+\Delta_1\tilde{\gamma}^2)x\nonumber\\
&&+(4\Delta^2_1+4\Delta_1\tilde{\gamma}+\tilde{\gamma}^2+2\Delta^2_1\tilde{\gamma}^2).
\end{eqnarray}
\begin{figure} 
\begin{center}
\includegraphics[width=.5\textwidth]{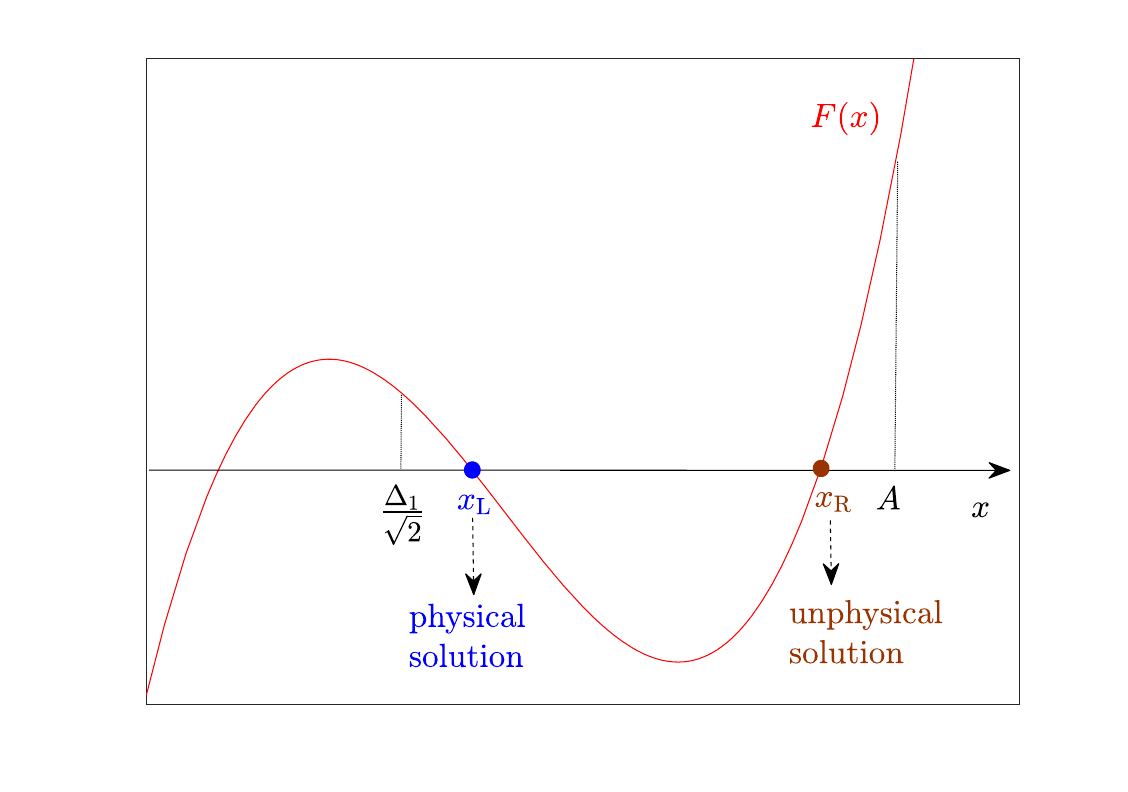}
\caption{Sketch of the graph of $F(x)$ given by Eq.~(\ref{Fx}). The equation $F(x)=0$ has a negative solution and two positive solutions on $x\in (\Delta_1/\sqrt{2},A)$. However, only one of these three solutions is physical and approximates $\cosh\tilde{p}$, where $\tilde{p}$ is the solution of Eq.~(\ref{tanh_gp}) at large $N$.}
\label{Fxgraph}
\end{center}
\end{figure}
From
\begin{eqnarray}
F(0)&=&4\Delta^2_1+4\Delta_1\tilde{\gamma}+\tilde{\gamma}^2+2\Delta^2_1\tilde{\gamma}^2>0,\nonumber\\
F(\Delta_1/\sqrt{2})&=& \tilde{\gamma}^2>0,\nonumber\\
F(A)&=& \tilde{\gamma}^4/2>0,
\end{eqnarray}
and
\begin{eqnarray}
\frac{dF}{dx}|_{x=\Delta_1/\sqrt{2}}&=&-2\sqrt{2}\tilde{\gamma}(2+\Delta_1\tilde{\gamma})<0,\nonumber\\
\frac{dF}{dx}|_{x=A}&=&\sqrt{2}\tilde{\gamma}^2(2\Delta_1+2\tilde{\gamma})>0,
\end{eqnarray}
we obtain a sketch of the graph of $F(x)$, see Fig.~\ref{Fxgraph}. As a cubic equation, $F(x)=0$ has three real solutions, one of which is negative and the remaining two are on $(\Delta_1/\sqrt{2},A)$. However, only one of the two positive solutions on $(\Delta_1/\sqrt{2},A)$ is a physical solution~\cite{PRB2024_2}, i.e., $x\approx\cosh\tilde{p}$, where $\tilde{p}$ is the actual solution of Eq.~(\ref{tanh_gp}) for large $N$. From $w^{(+)}(\Delta_1/\sqrt{2})=-\tilde{\gamma}\Delta_1/4<0$ and $w^{(+)}(A)=\tilde{\gamma}^2/4>0$, we see that the smaller (large) solution $x_{\mathrm{L}}$ ($x_{\mathrm{R}}$) on $(\Delta_1/\sqrt{2},A)$ actually corresponds to a physical (an unphysical) solution with $w^{(+)}(x_{\mathrm{L}})<0$ [$w^{(+)}(x_{\mathrm{R}})>0$] or $g^{(+)}(\Delta_1,\tilde{\gamma},\cosh^{-1}x_{\mathrm{L}})\approx 1$ [$g^{(+)}(\Delta_1,\tilde{\gamma},\cosh^{-1}x_{\mathrm{R}})\approx -1$].
\par Since $\Delta_1/\sqrt{2}<x_{\mathrm{L}}<A$, we have $R>\sqrt{2}/\tilde{\gamma}$ in regions III and IV. For $\tilde{\gamma}< \sqrt{2}$ we have $R>1$, while for $\tilde{\gamma}\geq \sqrt{2}$ there may exist a curve defined by $R=1$ on which the solution is given by $x=1/2+\Delta_1/\sqrt{2}$. By inserting this solution into Eq.~(\ref{Fx0}) we get the equation for the curve $R=1$ (see the black dotted curve in Fig.~\ref{Fig2}),
\begin{eqnarray}\label{unphyR}
\Delta_1 = \frac{\tilde{\gamma}^2-\sqrt{2}\tilde{\gamma}+2}{\sqrt{2}\tilde{\gamma}(\tilde{\gamma}-\sqrt{2})}.
\end{eqnarray}
At first sight, the above curve will separate the two regions with $R<1$ or $R>1$. However, numerical tests indicate that the curve defined by Eq.~(\ref{unphyR}) always corresponds to the unphysical solution $x_{\mathrm{R}}$ on $(\Delta_1/\sqrt{2},A)$. Since $R$ must vary continuously as the parameters are changed, we always have $R>1$ in regions III and IV with $\tilde{\gamma}\geq \sqrt{2}$. 
\par For odd $m$, we have $R=-1/(2\cosh\tilde{p}+\sqrt{2}\Delta_1)$, which obviously lies between $-1$ and $0$, and hence the solution of Eq.~(\ref{tanh_gm}) always yields an NNN-Ex BS. In summary, by investigating both the position of the eigenenergy and the structure of the corresponding eigenstate (at large $N$ for even $m$) of a BS, we reveal the emergence of the two types of two-magnon BSs in the five regions (see the last column of Table 1). 
\par It would be helpful to compare the behaviors of the two-magnon BS formation in the $k=-\pi/2$ sector with those in the $k=-\pi$ sector~\cite{PRB2024_1}. It is shown in Ref.~\cite{PRB2024_1} that for $S=1/2$ and $\Delta_1>1$, there exists a high-lying NNN-Ex BS above the continuum in the $k=-\pi$ sector, irrespective of the values of $\Delta_2$ and $\gamma$. On the other hand, for any set of parameters there is always a low-lying NN-EX BS below the continuum. These behaviors are in sharp contrast to case of $k=-\pi/2$, where the interplay of $\Delta_1$ and $\gamma/\Delta_2$ are essential in the formation of BSs.
\par Finally, we would like to mention that in an early work by Ono, Mikado, and Oguchi the two-magnon BSs in a spin-1/2 isotropic $J_1$-$J_2$ chain with $\Delta_1=\Delta_2=1$ are studied by using a fermionization method~\cite{Ono}. In particular, an analytical treatment of the $k=-\pi/2$ mode in the thermodynamic limit $N\to\infty$ shows that a BS always exists below or above the continuum band as $\gamma>2$ or $\gamma<2$ (see Fig.~3 in Ref.~\cite{Ono}). This is exactly consistent with our general results presented in Fig.~\ref{Fig2} (see the horizontal dashed orange line defined by $\Delta_1=1$). It is worth mentioning that our method and results are valid for any $\Delta_1$ and $\Delta_2$ and any finite $N$. 
\section{Conclusions}\label{SecVI}
\par In this work, we solve the two-magnon problem in the $k=-\pi/2$ sector of a finite-size spin-1/2 anisotropic $J_1$-$J_2$ ring. The method we employ combines a set of exact two-magnon Bloch states and a plane-wave ansatz method. For any number of sites $N$ that is divisible by 8, we semianalytically solve the eigenvalue problem of the Bloch Hamiltonian by writing the eigenvector as a plane-wave form in which the pseudo-momentum $p$ is determined by a transcendental equation. 
\par By analyzing the structures of the real and complex solutions of the above equation, we build up a phase diagram in the $J_1/(|J_2|\Delta_2)$-$\Delta_1$ plane where several regions supporting either the NN-Ex or the NNN-Ex two-magnon BSs are analytically identified. The evolution of the BSs with varying parameters can be directly read off from the phase diagram. Our results in the isotropic case are consistent with an early work by Ono, Mikado, and Oguchi. From physical continuity, we expect that the obtained analytical results in the $k=-\pi/2$ sector can faithfully reflect the properties of two-magnon BSs around the middle of the half-Brillouin zone. Our method might be applied to more complicated (quasi-)one-dimensional spin chains such as the Heisenberg octahedral chain~\cite{Karlova}, etc. \\
\\
\noindent{\bf Acknowledgments:}
This work was supported by the Innovation Program for Quantum Science and Technology under Grant No. 2023ZD0300700 and by the National Key Research and Development Program of China under Grant No. 2021YFA1400803.

\end{document}